\documentclass[12pt,a4paper]{article}
\newcommand{\idx}[1]{\mbox{\scriptsize [}#1\mbox{\scriptsize ]}}
\newcommand{\Idx}[1]{\mbox{\small [}#1\mbox{\small ]}}
\newcommand{\hc}{{\scriptscriptstyle{+}}}
\newcommand{\pht}[1]{\mbox{\phantom{#1}}}
\newcommand{\pow}{\raisebox{0.4ex}{\small$\,\uparrow\,$}}

\newdimen{\pxh}
\newdimen{\pxv}
\newcount{\BinPicN}
\def\binpic#1{{%
\def\h{1}%
\def\L##1{\hbox{{\R0}##1}\nointerlineskip}%
\def\V##1{\vskip##1\pxh\relax}%
\def\R##1{\vrule width##1\pxh height\h\pxv depth0pt}%
\def\S##1{\hskip##1\pxh\relax}%
\def\H##1##2{{\def\h{##1}{##2}}}%
\def\N##1##2{{\BinPicN=\number##1\loop\ifnum\BinPicN>0{##2}\advance\BinPicN by-1\repeat}}%
\def\0{\S4}\def\1{\S3\R1}\def\2{\S2\R1\S1}\def\3{\S2\R2}%
\def\4{\S1\R1\S2}\def\5{\S1\R1\S1\R1}\def\6{\S1\R2\S1}\def\7{\S1\R3}%
\def\8{\R1\S3}\def\9{\R1\S2\R1}\def\A{\R1\S1\R1\S1}\def\B{\R1\S1\R2}%
\def\C{\R2\S2}\def\D{\R2\S1\R1}\def\E{\R3\S1}\def\F{\R4}%
\hbox{\vbox{#1}}%
}}%
\def\picscale#1#2#3{{%
\ifx#1?\relax\else\pxh=#1\fi%
\ifx#2?\relax\else\ifx#2*\pxv=\pxh\else\pxv=#2\fi\fi%
#3%
}}%

\begin{document}
\pagestyle{empty}
\title{\bf\Large Von Neumann Quantum Logic {\em vs.}
Classical von Neumann Architecture?}
\author{{\sc A. Yu. Vlasov}\\
Federal Radiological Center, IRH\\
Mira Street 8, St.-Petersburg, Russia}
\date{}
\twocolumn[
\smash{$\mbox{\vtop{\vskip 4cm\relax{\Large\em Conference}\\
\mbox{\hskip 1pt\picscale{0.24bp}*{%
\binpic{%
\H3{\L{\0}}%
\L{\S{60}\1\4}%
\L{\S{28}\2\8\S{24}\2\A}%
\L{\S{28}\5\4\S{24}\5\5}%
\L{\S{28}\A\A\S{24}\A\A\8}%
\L{\S{24}\1\5\5\S{24}\5\5}%
\L{\S{24}\2\A\A\S{24}\A\A\8}%
\L{\S{24}\1\5\5\S{20}\1\5\5\4}%
\L{\S{24}\2\A\A\S{24}\A\A\8}%
\L{\S{24}\5\5\5\S{20}\1\5\5\4}%
\L{\S8\N2{\S{16}\2\A\A\8}}%
\L{\S{24}\5\5\5\S{20}\1\5\5\4}%
\L{\S{24}\2\A\A\8\S{16}\2\A\A\A}%
\L{\S{24}\5\5\5\4\S{16}\1\5\5\4}%
\L{\S{24}\A\A\A\8\S{16}\2\A\A\A}%
\L{\S{24}\5\5\5\4\S{16}\5\5\5\4}%
\L{\S{24}\A\A\A\8\S{16}\2\A\A\A}%
\L{\S{20}\1\5\5\5\4\S{16}\N4{\5}}%
\L{\S{24}\N4{\A}\S{16}\2\A\A\A}%
\L{\S{20}\1\5\5\5\4\S{16}\N4{\5}}%
\L{\S{24}\N4{\A}\S{16}\N4{\A}}%
\L{\S{20}\1\5\5\5\4\S{16}\N4{\5}}%
\L{\S{24}\N4{\A}\S{16}\N4{\A}}%
\L{\S{20}\1\5\5\5\4\S8\0\1\N4{\5}}%
\L{\S{20}\2\N4{\A}\S{16}\N4{\A}\8}%
\L{\S{20}\1\5\4\5\5\S8\0\1\5\4\5\5}%
\L{\S{20}\2\A\8\A\A\S{16}\A\A\2\A\8}%
\L{\S8\N2{\S{12}\1\5\4\5\5}}%
\L{\S{20}\2\A\8\A\A\S8\0\2\A\8\2\A\8}%
\L{\S{20}\5\5\4\5\5\S8\0\1\5\4\5\5\4}%
\L{\S{20}\N2{\2\A\8}\S8\N2{\2\A\8}}%
\L{\S{16}\7\F\D\0\5\5\S8\0\5\5\0\1\5\7\R8}%
\L{\S8\0\3\R8\F\C\3\E\8\S8\2\A\8\2\R{20}\0\7}%
\L{\S8\0\R{20}\7\D\4\S8\5\5\0\1\R{20}\C\F}%
\L{\S8\1\F\E\A\R8\B\E\8\S8\A\A\8\7\F\C\0\1\F\E\F}%
\L{\S8\3\F\0\5\5\R8\D\4\0\1\5\5\0\R8\4\S8\3\R8}%
\L{\S8\7\E\0\A\A\7\F\E\8\S8\A\A\3\F\E\8\S8\1\R8}%
\L{\S8\F\C\1\5\5\3\F\D\4\0\1\5\4\7\F\D\4\0\S8\R8}%
\L{\0\1\F\C\0\A\A\3\F\E\8\S8\A\A\7\F\A\A\0\S8\7\F}%
\L{\0\1\F\8\1\5\4\0\F\D\4\0\1\5\4\R8\5\4\0\S8\7\F}%
\L{\0\3\F\8\2\A\A\0\F\E\A\0\2\A\9\F\C\A\A\0\S8\3\F}%
\L{\0\3\F\8\1\5\4\0\7\D\4\0\1\5\7\F\D\5\5\0\S8\3\F}%
\L{\0\3\F\8\2\A\8\0\7\E\A\0\2\A\B\F\8\A\A\0\S8\1\F}%
\L{\0\7\F\8\5\5\4\0\7\D\5\0\5\5\7\F\8\5\5\S{16}\F}%
\L{\0\7\F\8\2\A\8\0\3\E\A\0\A\A\R8\0\A\A\S{16}\F\8}%
\L{\0\7\F\8\5\5\4\0\3\D\5\0\5\5\R8\0\5\5\S{16}\F\8}%
\L{\0\7\F\C\2\A\8\0\1\E\A\8\A\B\R8\0\A\A\8\S8\0\F\8}%
\L{\0\7\F\C\5\5\4\0\1\F\5\0\5\5\F\E\0\5\5\S{16}\7\8}%
\L{\0\3\F\C\A\A\8\0\1\E\A\8\A\B\F\E\0\2\A\8\S8\0\7\8}%
\L{\0\3\F\E\5\5\S8\0\F\5\1\5\5\F\E\0\5\5\S{16}\7\8}%
\L{\0\3\R8\A\A\8\S8\2\A\A\A\B\F\E\0\2\A\8\S8\0\3\8}%
\L{\0\1\R8\D\5\S8\0\N4{\5}\7\F\C\0\5\5\4}%
\L{\0\1\R8\F\A\S8\0\2\A\A\A\B\F\C\0\2\A\8}%
\L{\0\1\R{16}\S8\0\1\5\5\5\7\F\C\0\1\5\4}%
\L{\S8\R{16}\8\S8\2\A\A\A\R8\C\0\2\A\A}%
\L{\S8\7\F\R8\E\S8\1\5\5\5\7\F\C\0\1\5\4}%
\L{\S8\3\F\R8\F\8\S8\A\A\A\R8\C\0\2\A\A}%
\L{\S8\1\R{20}\0\1\5\5\5\7\F\C\0\1\5\5}%
\L{\S8\0\7\R{16}\8\0\A\A\A\R8\C\S8\A\A}%
\L{\S8\0\3\R{16}\E\1\5\5\5\7\F\C\0\1\5\5}%
\L{\S8\0\2\R{20}\0\A\A\A\7\F\C\S8\A\A}%
\L{\S8\0\1\5\F\R8\F\8\5\5\4\7\F\C\0\1\5\5}%
\L{\S8\0\2\A\B\R8\F\C\A\A\A\7\F\C\S8\A\A\8}%
\L{\S8\0\1\5\4\R8\F\E\5\5\4\7\F\C\S8\5\5}%
\L{\S8\0\2\A\8\1\R8\E\2\A\8\7\F\C\S8\A\A\8}%
\L{\S8\0\1\5\4\0\F\R8\1\5\4\7\F\C\S8\5\5\4}%
\L{\S8\0\2\A\8\0\7\R8\2\A\8\7\F\C\S8\2\A\8}%
\L{\S8\0\1\5\S8\3\R8\1\5\0\3\F\E\S8\1\5}%
\L{\0\E\S8\A\0\S8\R8\S8\0\3\F\E\S8\2\A}%
\L{\0\E\S{24}\R8\8\S8\3\F\E\S8\0\4}%
\H2{\L{\0\E\S{24}\7\F\8\S8\1\R8}}%
\L{\0\F\S{24}\7\F\8\S8\1\R8\S{32}\C}%
\L{\0\F\S{24}\7\F\8\0\S8\R8\8\S{24}\1\E}%
\L{\0\F\8\S{20}\7\F\S{16}\R8\8\S{24}\1\C}%
\L{\0\F\C\S{20}\7\F\S{16}\7\F\C\S{24}\3\C}%
\L{\0\F\C\S{20}\7\F\S{16}\7\F\E\S{24}\7\8}%
\L{\0\F\E\S{20}\7\E\S{16}\3\F\E\S{24}\7\8}%
\L{\0\F\E\S{20}\7\E\S{16}\1\R8\S{24}\F}%
\L{\0\R8\S{20}\7\E\S{16}\1\R8\8\S{16}\1\F}%
\L{\0\R8\8\S{16}\F\C\S{20}\R8\C\S{16}\3\E}%
\L{\0\R8\C\S{16}\F\8\S{20}\7\R8\S{16}\7\C}%
\L{\0\F\R8\S8\0\3\F\S{24}\3\R8\C\S8\3\F}%
\L{\0\F\R8\8\S8\7\E\S{28}\R8\F\8\1\R8}%
\L{\0\F\3\R8\0\3\F\C\S{28}\7\R{20}\C}%
\L{\0\F\1\R{20}\8\S{28}\1\R{20}\8}%
\L{\0\E\0\3\R8\F\C\S{32}\0\7\R8\F\C}%
\L{\S{16}\7\F\E\0\S{40}\R8\E}%
}%
}}\mbox{\huge$\!${\em '2000}} }}$}
\maketitle
\thispagestyle{empty}
{\small
\begin{quote}
{\em Abstract. }
The name of John von Neumann is common both in quantum mechanics and
computer science. Are they really two absolutely unconnected areas?
Many works devoted to quantum computations and communications are
serious argument to suggest about existence of such a relation, but it
is impossible to touch the new and active theme in a short review.
In the paper are described the structures and models of linear algebra
and just due to their generality it is possible to use universal description
of very different areas as quantum mechanics and theory of Bayesian
image analysis, associative memory, neural networks, fuzzy logic.
\end{quote}
}
\medskip
]
\sloppy

J. von Neumann is considered as one of ``fathers'' of modern
computers due to his theoretical works and elaboration of computer
EDVAC at 1945--50, but already 20 year before it, at \mbox{1926--31} he
participated in drawing yet another kind of logic --- logic of quantum
mechanics and only now the different areas of research of one
scientist seem going to meet and born new family of ultrascale
cybernetic devices --- {\em quantum computers}.

Currently the {\em quantum information science} exists mainly as
theoretical area of research and size of quantum registers does not
exceed of 3--5 quantum bits ({\em qubits}), but in the paper is
considered question: does the quantum logic has some useful
application as {\em an abstract mathematical model} in computer
science? Such applications of physical models nowadays are not
unusual, for example Boltzmann machines and other methods came
to area of artificial neural networks from statistical physics [1, 2].

It is also useful to draw some analogy with fuzzy sets and logic [3].
Here is used discrete representation of 2D sets on some lattice due to
understanding analogy with bit-map pictures.

Let us introduce few data types in some ``Pascal-like'' notation to
come from fuzzy sets to quantum registers. Here $n$ and $N = 2^n$
are \mbox{\bf integer} numbers.

1)      $x$ : {\bf dot} = [1 .. N, 1 .. N];

Here $x$ is represented as point on 2D space for simplicity,
because of digitization we can consider space of any dimension as
interval of natural numbers $x$~:~1~..~$N^2$. In this
example $x$ is visual model for $n + n = 2n$-bits register.

2) $I$ : {\bf set\_2D} = {\bf set of dot};

or

\pht{2)} $I$ : {\bf set\_2D} = {\bf array} [{\bf dot}] {\bf of Boolean};

The set, ``image'' can be considered as black and white picture --- black
set on white sheet. It is necessary $2^{N\cdot N}$ {\em i.e.}
$2\pow(2\pow(2n))$ bits for the set.

3)  $F$ : {\bf fuzzy$\!$\_set} = {\bf array} [{\bf dot}] {\bf of real};

The {\bf real} is interval [0.0 .. 1.0] of real numbers represented with
some finite precision $d$. It can be considered as an analogue of
gray-scaled picture. The fuzzy set also can be {\em standardized} by
condition $\sum_x{F\idx{x}} = 1$ where $\sum_x$ in example under consideration
is double sum:
\begin{equation}
\sum_{i,j=1}^N{F_{ij}} = 1
\end{equation}
(in continuous case an integration used instead of the summation and
distributions $F(x)$ are used instead of arrays).

The condition (1) is used in statistical interpretation of fuzzy set,
when it is suggested to choose some point of set with probability
proportional of {\em adequacy function} $F\idx{x}$, and sum
of all the probabilities is unit. Let us use notation $p\idx{x}$
for the probabilities.

\medskip

Now let us introduce notion of {\em quantum set }(it is not standard
term, mathematical object discussed further corresponds to
$2n$-{\em qubit register} in quantum information science [6] or quantum
mechanical system with $N^2$ states [7]):

4)  $Q$ : {\bf qu\_set} = {\bf array} [{\bf dot}] {\bf of complex};

Where {\bf complex} is $q = u + i\,w$, $|q| \le 1$.
Instead of standardization condition here is used {\em normalization}:
$\sum_x{|Q\idx{x}|^2} = 1$ {\em i.e.}:
\begin{equation}
\sum_{i,j=1}^N{\left|Q_{ij}\right|^2} = 1
\end{equation}
for example under consideration.

In quantum mechanics the complex numbers $q\idx{x}$ is
called {\em amplitudes}, related with classical probabilities as
$p= |q|^2 = u^2 + w^2$, {\em i.e.} there
is {\em standardized}  fuzzy set related with given  {\em normalized}
quantum set via formula: $p\idx{x} = |q\idx{x}|^2$.

It is possible for simplification to consider case with real amplitudes
($w~=~0$), and let us explain why an ``auxiliary'' fuzzy set with
$q\idx{x}=\sqrt{p\idx{x}}$ has some independent useful application.

Let us consider two sets $p_1\idx{x}$, $p_2\idx{x}$ and look for some
{\em likelihood } function $H(p_1, p_2)$ with properties:
$H(p_1,p_2) < 1$ for $p_1\ne p_2$, $H(p,p) = 1$ [1, 3]. For standardized
sets such function can be
chosen as $H(p_1,p_2) = \sum_x{(p_1\idx{x} p_2\idx{x})}^{1/2}$. If
we use quantum sets $q_1$, $q_2$ ($p_1=q_1^2$, $p_2=q_2^2$),
the formula is $H(q_1, q_2) = \sum_x{q_1\idx{x} q_2\idx{x}}$.

In our example $x$ was chosen as multi-index of 2D array mostly
for simple visualization and any array can be described as one-dimensional.
Here $x = (i, j)$ can be substituted by
index of the 1D array like $K~=~(i-1)N+j$,
$K \in [1\ ..\ N^2]$, then formula can be written
as:
\begin{equation}
H(q,q')=\sum_{i,j=1}^N{q_{ij} q'_{ij}} \equiv \sum_{K=1}^{N^2}{q^{\,}_K q'_K}
\end{equation}
and it shows, that $H(q,q')$ is simply scalar
product of two vectors with $N^2$ elements and unit lengths
(due to normalization condition) and so the function really can be
unit only for equivalent vectors.

It should be mentioned also, that the formula (3) is given for {\em
real} $q^{\,}_K$ for simplicity and in {\em complex} case it contains
terms with complex conjugation: $q^{\,}_K\bar{q}'_{K}$. It is {\em Hermitian}
norm [7]. Such case also has applications, for example then we work with
{\em complex Fourier} {\em transform} of some image.

The discussed property of quantum set as ``square root of fuzzy set''
makes clear, why it can be useful in such abstract area, as image
recognition [4, 5], quite far from initial appearance in quantum
mechanics.

Let us discuss now some question related with  ``hardware''. We had
few data structures: \\
1) $x$ : {\bf dot}, 2) $I\idx{x}$ : {\bf set\_2D},
3) $p\idx{x}$ : {\bf  fuzzy$\!$\_set}\\
It is sequence with more and more complicated structure with
occupation of more and more computer memory. But let us suggest,
that register $x$ is permanently changing its value by such a law,
that after enough period of time $T$ it can be found, the $x$
had value $v$ during time $t\idx{v}$ and
$\lim\limits_{T\to\infty}t\idx{v}/T=p\idx{v}$.

The algorithm can be implemented by software, but it also can be
considered as some hardware register (like implementation of random
number generator in some computers with main difference, that
$p\idx{v}$ depends on $v$; or input port of some analog-to-digital
converter is scanning some ``physical model'' of fuzzy set $p$).

Then each access to the register produces some value of $x$ with
probability $p\idx{x}$,  so one $2n$-bits stochastic register
is enough to implement statistical model of standardized fuzzy set
discussed earlier.

\medskip

Now let us come to {\bf qu\_set} data type. Why it can be modeled by
one quantum register? The example with stochastic register as model
of fuzzy set is some analogy (see also [8]). But procedure of access to
such register due to laws of quantum mechanics has some differences
with classical statistical register discussed below.

Let us consider some value $q \in$ {\bf qu\_set} of the register, it is
array of numbers $q\idx{x}$, we may not read all the numbers,
but if we access to the register, we read number $x$ with
probability $p\idx{x} = |q\idx{x}|^2$ and it
coincides with functionality of stochastic register described above.

It should be mentioned only, that any access to $q$ destroys the
quantum register by substitution instead of $q$ new array with 1
in element with index $x$ and with all other is 0, and so the
register should be reset ({\em preparation} in terminology of
quantum mechanics) in $q$ after each access (quantum {\em
measurement}).

But the quantum register has other useful property, it is possible
instead of simple access described above to perform another
operation, we prepare some given $q' \in$ {\bf qu\_set} and read
the register $q$ with using $q'$ as some ``quantum bit-mask'',
then with probability $|H(q,q')|^2$ (see
Eq. 3) operation is successful and so by repeating it more times we
may found $|H|$ with more precision.

Because the $|H|$ has useful application as likelihood function,
the quantum register can be used as some hardware accelerator for
image analysis. Currently such hardware is not accessible and so it
was interesting to research advantages and disadvantages of the
particular function $|H|$ in usual software applications.

It is promising not only because such kind of software would suffer
giant speed-up after creation specific quantum hardware, but also
because the used mathematical constructions and methods of linear
algebra are quite convenient and powerful.

It should be mentioned, that similar mathematical methods already
was used in models of associative memory [9] and formal neural
network [10] without any relation with quantum mechanical models.
Only noticeable difference was using {\em real} linear spaces instead
of {\em complex} and Euclidean norm (Eq. 3) instead of Hermitian
(with complex conjugation).

\medskip

Let us now consider  some operations with fuzzy and quantum sets,
discuss fuzzy and quantum {\em logic}.

For usual sets we have basic operations for $A$,~$B$: {\bf set\_2D},
 --- {\em intersection}: $A\cap B$, {\em union}: $A\cup B$, {\em
complementation}: $A^c$. With using presentation of set as
Boolean array the operations in components can be written:
$(a^c)\idx{x}~=~\mbox{\bf not}~a\idx{x}$, $(a\cap b)\idx{x} =  a\idx{x}\land b\idx{x}$,
$(a\cup b)\idx{x} = a\idx{x} \lor b\idx{x}$.

Similarly it is possible to define operations with fuzzy sets by
definition of {\em real }analogs of Boolean operations {\bf not }($\lnot$),
{\bf and} ($\land$), {\bf or} ($\lor$).

For example $\lnot a \mapsto 1-a$, $a \land b \mapsto \min(a,b)$,
$a \lor b \mapsto \max(a,b)$ is a good choice, but here is used a second one,
more algebraic $a\land b \mapsto a\cdot b$,
$a \lor b \mapsto a+b-ab = \lnot ((\lnot a) \land (\lnot b))$.

But {\bf qu\_set} introduced above is not directly used in quantum
logic --- the linear operators are used here instead of vectors:

$L$ : {\bf qu\_map} = {\bf array} [{\bf dot},{\bf dot}] {\bf of complex};\\
It is matrix for linear map: {\bf qu\_set} $\to$ {\bf qu\_set}:
\begin{equation}
q'_{ij}= \sum_{k,l=1}^N{L^{\,}_{ij,kl} q'_{kl}} \quad \mbox{or} \quad
q'_I = \sum_{K=1}^{N^2}{L^{\,}_{IK} q'_K}
\end{equation}
where indexes $I$, $K$ are used instead of multi-indexes
\idx{$i,j$} and \idx{$k,l$} (let us for simplicity use
further the indexes like $I, K : [1\ ..\ M]$, $M=N^2$).

The operators $A, B,\ldots \in$ {\bf qu\_map} form an
algebra with usual matrix multiplication {\em C=AB}:

\begin{equation}
C_{KJ}= \sum_{I=1}^M{A_{KI} B_{IJ}}
\end{equation}

A special kind of operators, {\em projectors}, make possible
comparison of the algebra with logic, {\em i.e.} Boolean algebra. The
projector is operator with property $P^2 = P$. Let us
consider set of orthogonal projectors, {\em i.e.}
$P_i P_j = 0$, $i\ne j$, then the operators produce
Boolean algebra in respect of operations:
\begin{equation}
\lnot P{\equiv}1{-}P,\, P{\land}R{\equiv}PR,\, P{\lor}R{\equiv}P{+}R{-}PR  
\end{equation}

Elements of the algebra have form $P_{(S)}$ there $S$~:~{\bf set of}~1~..~$M$:
\begin{equation}
P_{(S)}= \sum_{I \in (S)} {P_I}
\end{equation}

There is relation between $q\in$ {\bf qu\_set} and some projector
$P_q{\in}${\bf qu\_map}: $P_q\Idx{I,J}=q\Idx{I}q\Idx{J}$.
To describe properties of the
projector it is convenient together with $q$ considered as {\em
row} with $M$ elements to consider transposed {\em column}
$q^\hc$ (conjugated for complex case).

Then $P_q = q^\hc \cdot q$ and $q\cdot q^\hc = |H(q,q)|^2~=~1$ (the row
$q$ can be considered as $1{\times}M$ matrix, column $q^\hc$ as
$M{\times}1$ matrix and due to law of
multiplication the $q\cdot q^\hc$ is $1{\times1}$ matrix {\em
i.e.} number and $q^\hc\cdot q$ is $M{\times}M$ matrix) and
\mbox{$P_q P_q = q^\hc\cdot q\cdot q^\hc\cdot q = q^\hc\cdot1\cdot q =
q^\hc\cdot q = P_q$}.

If we consider family of nonintersected sets $q_1$, $q_2, \ldots, q_k$
then $P_i = q_i^\hc\cdot q_i$ are orthogonal projectors and so
quantum sets in such representations have rather relations with usual
logic than with fuzzy one.

For more clear explanation of properties of $P_i$ it is
possible to use existence of some orthogonal (unitary for complex
case) matrix $U$ same for all $P_i$ such, that all
$P'_i = U P_i U^{-1}$ are diagonal and have
very simple form: $P'_1= \mbox{diag}(1,0,$\ldots$,0)$,
$P'_2 = \mbox{diag}(0,1,\ldots,0)$, {\em etc.}.

The classical Boolean structure of the operators $P_i$ and
their sums $P_{(S)}$ (see Eq. 7) is because of all the
operators commute [11]. If to choose projectors $P$, $R$:
$PR\ne RP$ the Eq. 6 do not produce Boolean algebra, but it
is other kind of non-Boolean logic, than fuzzy one.

It should be mentioned, that more direct relation with fuzzy set have
so-called {\em mixed} quantum states $R = \sum_i{w_i P_i}$,
$\sum_i{w_i} = 1$ where $w_i$ have statistical nature and so here is
written analog of standardized fuzzy set.

A representation of some kind of fuzzy operations is example with
family of commuting operators [12], but not projectors. They are
described by diagonal matrices: $D_q\Idx{I,I} = q\Idx{I}$,
$D_q\Idx{I,J} = 0$, $I\ne J$ and already shown Eq. 6.

\smallskip

{\em Bibliographical notes}: In addition to references [7, 11] some
general handbook on quantum mechanics like [13] is appropriate for
most text of the paper. New area of quantum computation is
presented in [6, 14--17]. Two works
of J. von Neumann devoted to quantum logic [18] and electronic
computers [19] are included for completeness.

{\footnotesize
\renewcommand{\refname}{{\hfil\bf\normalsize References\hfil}}

}
\end{document}